\documentclass[12pt]{article}
\usepackage{graphicx}
\usepackage{amsmath}

\newcommand{\rcorr}{\hbox{\kern-1.2em$\longrightarrow$}}
\newcommand{\lrcorr}{\hbox{\kern-1.2em$\longleftrightarrow$}}

\newcommand{\nop}{\hbox{{\textsf I}\kern-.47em\hbox{O}}}

\def\TREV{{{}^\triangleleft\kern-1.5pt\texttt{T}}}
\def\trev{{{}^\triangleleft\kern-3.2pt\texttt{t}}}
\def\SREV{{{}_\triangleleft\kern-2pt\texttt{S}}}
\def\srev{{{}_\triangleleft\kern-2.2pt\texttt{s}}}
\newcommand{\Id}{\hbox{\sl 1\kern-0.25em\hbox{I}}}

\newcommand{\nRightarrow}{\Rightarrow\kern-1.2em\hbox{/}\kern.8em} %
\newcommand{\BB}{\hbox{I\kern-.2em\hbox{B}}} 
\newcommand{\DD}{\hbox{I\kern-.2em\hbox{D}}} 
\newcommand{\FF}{\hbox{I\kern-.2em\hbox{F}}} 
\newcommand{\NN}{\hbox{I\kern-.2em\hbox{N}}}  
\newcommand{\ZZ}{{{\rm Z}\kern-.28em{\rm Z}}} 
\newcommand{\RR}{\mathop{{\rm I}\kern-.2em{\rm R}}\nolimits} 
\newcommand{\QQ}{\hbox{l\kern-.36em\hbox{Q}}}  
\newcommand{\CC}{\hbox{I\kern-.58em\hbox{C}}}

\begin{document}
\title{\bf Consistent Value Assignments\\ Can Explain Classicality}
\author{Giuseppe Nistic\`o\\
{\small Dipartimento
di Matematica e Informatica, Universit\`a della Calabria, Italy}\\
{\small and}\\
{\small
INFN -- gruppo collegato di Cosenza, Italy}\\
{\small address: Via Bucci 30B, 87036 Rende (taly); phone 0039(0)984496413}\\
{\small email: giuseppe.nistico@unical.it}. ORCID: 0000-0002-9515-4015 }
\maketitle
\abstract{The present work proposes an alternative approach to the problem of 
the emergence of classicality.
Typical approaches developed in the literature derive the classical behaviour of 
a quantum system from conditions that concern the value of the
parameters deemed responsible of non-classicality, 
like Planck constant.
\par
Our first step in addressing the problem is instead to identify the physical origin
of non-classicality of quantum physics. Nowadays the deepest origin
is identified in the impossibility of a simultaneous consistent value assignment 
to every set of quantum observables. To attack this impossibility a concept 
of ``evaluation'' is then introduced, which allows for a consistent value 
assignment to non-comeasurable observables whenever an established
set of conditions is satisfied. It is shown that in the case of the motion of the center of mass 
of a large rigid body evaluations exist that realize a consistent value assignment
to both the position and the velocity of the center of mass of the body.
In so doing emergence of classicality is explained by overcoming the
obstacles to the simultaneous value assignments that allow for a classical 
description of the phenomenon. This result prompts to search for 
extensions and generalization of the approach.
}
\section{Introduction}
In the early twentieth century empirical failures of classical physical theories made necessary 
to replace them with  new theories, quantum theories, empirically valid.
One problem of these new theories is emergence of classicality, in rough synthesis the problem of explaining
why at a macroscopic level world is classical -- more generally, why
quantum systems with ``classical'' features (e.g. being
``macroscopic'') behave following classical laws, rather than quantum laws.
The present work shows that the question can be addressed from a yet unexplored side, as we are going to explain.
\par
The very reason of the empirical failure of classical theories can be nowadays
identified within their deepest conceptual basis.
Let us be  synthetically more explicit. The development of every classical theory requires that each
specimen of the physical system is assigned a value $v({\mathcal A})$  for every physical
magnitude $\mathcal A$ of the physical system at every time, as an objective property of that specimen. In fact,
no classical theory can be developed without the following assumption.
\begin{itemize}
\item[($\mathcal{CS}$)]
{\sl At any time every specimen of the physical system has a determined 
value $v({\mathcal A})$ of each of its magnitudes $\mathcal A$.
An eventual measurement of $\mathcal A$, with outcome $v_0({\mathcal A})$, just
reveals this objective value.}
\end{itemize}
Now, in empirically ascertained phenomena {\sl any} simultaneous value assignment turned 
out to be empirically inconsistent, as shown in appendix A.
Therefore, alternative physical theories must be developed,
consistent with the following statement.
\begin{itemize}
\item[($\mathcal{QS}$)]
{\sl A specimen of the physical system is not always endowed
with a value of every magnitude as an objective property of the specimen.}
\end{itemize}
Quantum theories are consistent with ($\mathcal{QS}$); in fact they replaced classical theories.
The present work shows how just the limitations to a simultaneous consistent value assignment
can be outflanked according to quantum theory in some cases where the physical features of the system become 
``classical'': classicality emerges in virtue of a mechanism at the level of the deep roots of quantum physics. 
\par
This approach is different from the past ones we briefly outline.
The primary approach to the question involved the {\sl Planck constant} $\hbar$ (e.g. see \cite{Hagerdon},\cite{Robert}); it aimed 
to formally identify the conditions, within the specific quantum theory of the involved system, such that
in the limit $\hbar\to 0$ the physical system obeys classical laws.
Another approach 
is based on the idea of {\sl decoherence} (e.g. see \cite{Zurek}), according to which
the Hilbert space of the quantum theory must be enlarged to take into account the
(albeit very weak) interaction with the environment; the conditions  are then searched which imply
the ``collapse'' of the wave function, i.e. loss of the superposition of quantum states, with respect to a reduced  description that excludes the environment, so that
a classical behaviour emerges at this level.
A further idea pursued to solve the question is to investigate the consequences of the {\sl largeness}
of a system, i.e. of the fact that the system is composed by a very large number $N$ of identical
constituents \cite{Landsman};
one result of this approach is that
{\sl macro-observables}, i.e. observables of a macroscopic system which are ``symmetrized'' averages  of observables  
of the constiuents,
turn out to commute in the limit $N\to\infty$, so that the class of these macroscopic observables is isomorphic to a class of classical magnitudes.
\par
These different lines of research are treated in great detail in the literature
\footnote{A different interesting investigation has been pursued \cite{Teta}, 
showing how the
classical trajectories of a particle in a cloud chamber emerge without making
resort to the idea of collapse of the wave function, but, instead, by a pure quantum theoretical
treatment.}
\cite{Robert},\cite{Landsman}.
In particular Landsman \cite{Landsman}  provides us with a truly extended collection of references.
In general these approaches aim to identify under which conditions of the specific quantum theory classicality can be derived 
when the features that exhibit non-classicality vanish;
for instance, the primary approach investigates under which conditions the quantum theory of 
a system predicts a classical behaviour in the limit $\hbar\to 0$.
These features
are themselves consequences of the theoretical re-development that leads to quantum theory coherently with $({\mathcal QS}$).
As concluded by Landsman himself, ``none of these ideas in isolation is capable of 
explaining the classical world, but there is some hope that by combining [...] them, one might do so in the future''.
In our approach, instead, emergence of classicality stems from a mechanism that overcomes just the limitations to a simultaneous
consistent value assignment, which are the roots of non-classicality. In so doing, emergence of classicality is explained at a deeper
conceptual level.
\vskip.5pc
Let us describe the content of the work in more detail.
The first task accomplished in Section 2 is to show how quantum theory reformulates the basic physical concepts
to comply with ($\mathcal{QS}$). In particular, the classical concept of {\sl magnitude} is replaced by the
concept of  quantum {\sl observable} that does not necessitate a value assignment, unless it is actually measured.
The standard formulation of quantum theory in Hilbert space is presented,
where observables are identified with self-adjoint operators.
\par
Actually performed measurements of observables assign them the values obtained as outcomes, 
and therefore measurement determine value assignments empirical consistent.
In quantum theory a law that rules over a kind of limitation to simultaneous value assignments to different observables
is formally proved:
two obervables $A$ and $B$ that do not commute cannot be measured together on
the same specimen of the physical system.
However, this limitation to similtaneous value assignment by measurements does not exclude the
possibility of an otherwise obtained consistent value assignment to observables that are not co-measurable.
\par
In Section 2.2 the concept of {\sl evaluation} is introduced. Let an observable $A$ be assigned
the value obtained as outcome of a measurement of another observable $T$. 
A consistency criterion is formally established in such a way
that, when satisfied, the assigned value can be considered ``objective'' without the risk of 
logical contradiction or empirical inconsistency.
Whenever this criterion is satisfied the assignment is called {\sl evaluation} of $A$ by $T$.
In fact, evaluations are daily praticized in physics. 
For instance, the experimental determination of the spin of a silver atom by means of a Stern-Gerlach apparatus
is an evaluation, not a measurement. Indeed in this experiment the value of the spin is assigned according
to the exit, up or down, the atom goes out from the magnet; this procedure corresponds to a position measurement,
not to a spin measurement;
this position measurement turns out to be an evaluation of the spin.
\par
Evaluations are tools that can allow to overcome limitation to simultaneous value assignment to explain
emergence of classicality, according to the following arguments.
Let two observables $A$ and $B$ admit respective evaluations  by $T$ and $S$;  provided that $T$ and $S$ commute,
their simultaneous measurement give rise to the simultaneous evaluation of $A$ and $B$,
also in the case that $A$ and $B$ are not co-measurable.
In Section 3 this idea is exploited to explain the classicality of the motion
of the center of mass of a large ``rigid'' body.
The classical description of this motion
 requires the simultaneous value assignment to the position ${\bf Q}_{CM}$
and to the velocity ${\bf V}_{CM}$ of the center of mass. According to the quantum theory of this system 
$[{\bf Q}_{CM},{\bf V}_{CM}]\neq\nop$, therefore the simultaneous value assignment by direct  measurements
is forbidden. It is proved that just the condition of being rigid, formalized in the quantum theory, implies the existence
of an evaluation of ${\bf Q}_{CM}$ by ${\bf T}_{Q}$  and also of an evaluation of ${\bf V}_{CM}$ by ${\bf T}_{V}$,
such that $[{\bf T}_{Q},{\bf T}_{V}]=\nop$. Therefore both ${\bf Q}_{CM}$ and ${\bf V}_{CM}$ can be simultaneously
assigned consistent value obtained as outcomes of a simultaneous measurement of ${\bf T}_{Q}$ and ${\bf T}_{V}$.
\par
The success in explaining emergence of classicality for a ``rigid body'' as here defined, makes viable
the possibility of extensions and generalization of the approach.
\section{Quantum theory and value assignment}
Every classical physical theory assumes that each specimen of the physical system is assigned a value
for every  physical magnitude, as an objective property of that specimen, i.e. assumes ($\mathcal{CS}$).
The discovery of real experiments (see Appendix A) that contradict the predictions implied
by this assumption destroys the validity of this
 picture\footnote{In fact the experiments prove inconsistency of statement ($\mathcal{CS}$)
with empirically validated relations between particular magnitudes and the
empirically validated principle of {\sl locality}.}.
Therefore, to attain an empirically valid theory a
reformulation of the theories  is necessary, coherent with the following re-fundative statement.
\begin{itemize}
\item[($\mathcal{QS}$)]
{\sl A specimen of the physical system is not always endowed
with a value of every magnitude as an objective property of the specimen.}
\end{itemize}
The results are quantum theories.
The failure of assumption ($\mathcal{CS}$) entails that also the basic physical concepts must be re-defined.
Quantum theory replaces the classical concept of physical magnitude with the concept of
{\sl observable}. An observable $\textsf A$ identifies a process,
a {\sl concrete procedure}, which whenever applied to individual concrete specimens
of the physical system yields each time a real number $a$;
such a process is called measurement procedure, or simply {\sl measurement}, and $a\equiv v_0({\textsf A})$ is called
{\sl outcome} of that measurement of $\textsf A$.
This concept of observable does not require
to assign each specimen of the physical system with a value $v({\textsf A})$ for any observable $\textsf  A$
if not measured, and therefore it is coherent with ($\mathcal{QS}$).
\par
The task of a theory coherent with ($\mathcal{QS}$) cannot longer be to establish the
relations among the classically alleged ``objective'' values of the magnitudes, but it becomes
to establish the relationships among the {\sl actual occurrences} of physical phenomena.
These occurrences include the occurrences of measurements' outcomes,
and any phenomenon itself can be identified with one or more suitable occurrences of measurements' outcomes.
Therefore, quantum theory accomplishes its task by establishing the relationships among 
actual occurrences of outcomes of measurements of observables.
\par
Subsection 2.1 shows how every quantum theory, in its Hilbert space formulation \cite{VonNeumann},
is able to encompass the new concept of {\sl observables} coherently with ($\mathcal{QS}$),
by identifying themwith the self-adjoint operators of the underlying Hilbert space.
One important standard result related to value assignment establishes that if two operators
 do not commute then the corresponding observables cannot be measured together.
\par
Subsections 2.2 and 2.3 introduce and investigate the concept of consistent value assignment within quantm theory. 
It is shown how and under which conditions
also observables that cannot be measured together can admit
an empirically consistent similtaneous value assignement whenever determined conditions are realized.
This is done by introducing the concept of {\sl evaluation}. 
\subsection{Basic Quantum Theory}
The formally and conceptually coherent formulation of quantum theory 
of a specific physical system is grounded on a Hilbert
space $\mathcal H$, complex and separable.
Within the Hilbert space realization of the quantum theory of a specific system,
each observable $\textsf A$ is represented by a self-adjoint operator $A$,
whose set is denoted by $\Omega({\mathcal H})$.
We shall identify an observable $\textsf A$ with the self-adjoint operator $A$ that represents it,
whenever no confusion is likely.
The set $\tilde\sigma({\textsf A})\equiv \tilde\sigma({A})$ of the possible 
outcomes of measurements of $\textsf A\equiv A$ is called
physical spectrum of $A$.
\vskip.5pc
Now we recall some well known results of the development of quantum theory of interest for the present work.
\vskip.5pc\noindent
 {\sl Expected value.}
The expected value of the outcome of a measurement of $A$ is
$$
\textsf{Ev}({\textsf A})\equiv \textsf{Ev}({ A})=Tr(\rho A),$$
where $\rho$ is the {\sl density operator} that
represents the {\sl quantum state} of the system \cite{VonNeumann}.
As an implication, the physical spectrum $\tilde\sigma({A})$ turns out to coincide with the
mathematical spectrum $\sigma(A)$ of the corresponding operator $A$.
\vskip.4pc\noindent
{\sl Elementary observable.}
By {\sl elementary} observable we mean any observable $E$
having only $0$ or $1$ as possible outcomes, i.e. such that $\tilde\sigma(E)\subseteq\{0,1\}$.
As a consequence,
the expected value of an elementary observable $E$ must coincide with the
{\sl probability} $p_\rho(E)$ that in a measurement of $E$ the outcome $e=1$ occurs:
$$p_\rho(E)=\textsf{Ev}(E)=Tr(\rho E).$$
The coincidence $\sigma(E)=\tilde\sigma(E)$ implies that $E$ must be a
projection operator, and therefore the set of all elementary observables
must be identified with the set
$\Pi({\mathcal H})$ of all projection operators.
\par
If $E$ is an elementary observable, by $E'$ we denote the elementary observable whose outcome is $1$
(resp., $0$) if the outcome of $E$ is $0$ (resp., $1$). Accordingly, $E'=\Id-E$.
\par
According to spectral theory \cite{Kreyszig},
every self-adjoint operator $A$ admits the spectral representation
$A=\int \lambda\,d\,E_\lambda$ where $E:\RR\to \Pi({\mathcal H})$, $\lambda\to E_\lambda$,
is the resolution of the identity of $A$
and the integral converges with respect to the weak
convergence criterion of $\mathcal H$.
Hence the expected value of $A$ is $\textsf{Ev}({A})=Tr(\rho A)=\int\lambda\,d\,Tr(\rho E_\lambda)$, so that
$\int_{\lambda_1}^{\lambda_2}\hskip1pt d\hskip1pt Tr(\rho E^A_\lambda)=Tr(\rho(E^A_{\lambda_2}-E^A_{\lambda_1}))$
is just the probability $p_\rho((E^A_{\lambda_2}-E^A_{\lambda_1})$
that a measurement of $A$ yields an outcome in $(\lambda_1,\lambda_2]$, if the quantum state of the system is $\rho$.
\vskip.4pc\noindent
{\sl Functional principle.}
Given a self-adjoint operator $A$ and a real function $f$, according to spectral functional calculus
\cite{Kreyszig} another operator $f(A)$ can be defined as $f(A)=\int f(\lambda){\hskip1pt}d{\hskip1pt}E^A_\lambda$.
Hence the equalities
\par\noindent
$
\;\textsf{Ev}(f(A))=Tr(\rho f(A))=\int f(\lambda)d\hskip1pt Tr(\rho E^A_\lambda)=
\int f(\lambda){\hskip1pt}d{\hskip1pt}\textsf{Ev}(E^A_\lambda)=
\int f(\lambda){\hskip1pt}p_\rho(E^A_{\lambda+d\lambda}-E^A_\lambda)
$
entail that a measurement of the quantum observable $f(A)$ can be performed
by measuring $A$ and then transforming the outcome $a$ by the function $f$
into the outcome $b=f(a)$ of $f(A)$.
\vskip.4pc\noindent
{\sl Co-measurability and value assignment.}
If two self-adjoint operators $A$ and $B$ commute, then according to functional analysis
a third self-adjoint operator
$C$ and two functions $f$ and $g$ exist such that $A=f(C)$ and $B=g(C)$ \cite{Kreyszig}.
Therefore $A$ and $B$
can be measured together on the same specimen by measuring $C$, so that the outcome $a$ of $A$ and
$b$ of $B$ can both be obtained as $a=f(c)$ and $b=g(c)$, where $c$ is the actually measured value of $C$.
\par
Conversely, if $A$ and $B$ can be measured together on the same specimen
in all quantum states, then a third observable $C$ and two functions $f$ and $g$ can be identified
such that $A=f(C)$, $B=g(C)$ and thus $[A,B]=[f(C),g(C)]=\nop$. Indeed, any simultaneous measurement
of $A$ and $B$ yields a pair $(a,b)$ of outcomes; the outcomes of $A$ and $B$ can be recovered
by the functions $\alpha$ and $\beta$ defined by $\alpha(a,b)=a$ and $\beta(a,b)=b$ on the set $\{(a,b)\}$ of all
possible such pairs. A bijective correspondence, in general non continuous,
$\gamma:\{(a,b)\}\to \Gamma\subset\RR$ exists. Then every simultaneous measurement of $A$ and $B$
identifies a value $c=\gamma(a,b)$, so that a third observable $C$ can be defined whose
measurement is performable by measuring $A$ and $B$ together and whose outcome is just
$c=\gamma(a,b)$. Accordingly, $a=\alpha(\gamma^{-1}(c))$ and $b=\beta(\gamma^{-1}(c))$.
Therefore, $A=f(C)$ and $B=g(C)$, where $f=\alpha\circ\gamma^{-1}$ and $g=\beta\circ\gamma^{-1}$,
and then $[A,B]=\nop$. Thus the following algebraic characterization holds.
\vskip.5pc\noindent
{\bf Theorem 2.1.}
{\sl $A$ and $B$ are measurable together if and only if $\;[A,B]=\nop$.}
\vskip.5pc\noindent
{\bf Remark 2.1.}
Theorem 2.1 implies that given $E,F\in\Pi({\mathcal H})$, if $[F,E]=\nop$, the probability that
in a simultaneous measurements both outcomes turn out to be 1 is $Tr(\rho EF)$.
As a consequence, whenever $E,F\in\Pi({\mathcal H})$ are orthogonal, i.e. if $EF=\nop$ or equivalently
if $E+F\in\Pi({\mathcal H})$, the outome 1 for $E$ excludes the oucome 1 for $F$ in every simultaneous measurement.
\vskip.4pc\noindent
If $A$, $B$, $...$ is a set of observables actually measured together on the same specimen, then
to assign them
the respective outcomes $a$, $b$, $...$
realizes a value assignment consistent with all actually occurring phenomena,
because the occurrences of measurement outcomes
are actually occurring phenomena, and therefore no contradiction can take place with the other actually occurring phenomena.
Hence, co-measurability of $A$, $B$, $...$ implies the existence of consistent value assignment.
This implication is equivalent to the statement
\vskip.5pc\noindent
{\bf Proposition 2.2.}
{\sl Whenever observables $A$, $B$, $...$ 
do not admit a simultaneous consistent value assignment,
then they cannot be measured together.}

\subsection{Evaluations}

By itself, however, Prop.2.2 does not exclude existence of a consistent value assignment
for non co-measurable observables $A$, $B$, $...$.
In fact, under particular circumstances also non-measurable together observables
can be assigned values by means of ``evaluations'' introduced in \cite{Ev}, we are going to define after having premised
Def. 2.1 and Theorem 2.2.
\vskip.5pc\noindent
{\bf Definition 2.1.}
{\sl Two quantum observables $A$, $T$ are said to be perfectly correlated
with respect to the quantum state $\rho$ if
\begin{itemize}
\item[i)]
they are co-measurable,
\item[ii)]
 in every simultaneous measurement
the two outcomes coincide.
\end{itemize}}
\noindent
Perfect correlations are mathematically characterized by the following theorem whose proof is postponed in appendix B.
\vskip.5pc\noindent
{\bf Theorem 2.2.}
{\sl Two quantum observables $A$ and $T$ are perfectly correlated with respect to the quantum state $\rho$,
in formula $A\quad^\rho\lrcorr T$, if and only if
\begin{itemize}
\item[i)]
$[A,T]=\nop$,
\item[ii)]
$A\rho=T\rho$ or $\rho T=\rho A$.
\end{itemize}
If $\rho$ is a pure quantum state, i.e. if $\rho=\vert\psi\rangle\langle\psi\vert$, 
(ii) is equivalent to $A\psi=T\psi$.
}
\vskip.5pc\noindent
{\bf Corollary 2.2.}
{\sl 
If $[A,T]=\nop$, then $A\rho=T\rho$ if and only if $\rho T=\rho A$.}
\vskip.5pc
Now we can introduce {\sl evaluations}.
\vskip.5pc\noindent
{\bf Definition 2.2. (Evaluation)}
{\sl
Whenever $A\quad^\rho\lrcorr T$ holds, the value assignment to $A$ of the outcome of $T$
is called evaluation of $A$ by $T$.}
\vskip.5pc\noindent{\bf Example 2.1.}
An effective example of evaluation is the assignment of the
value of the spin observable to a silver atom through a Stern and Gerlach
apparatus, where the gradient of the magnetic field of the magnet is oriented along $z$ (Fig. 1).
Here we identify with 1 (resp., 0) the outcome $+1/2\hbar$ (resp., $-1/2\hbar$).
The $z$ component $S_z$ of the spin is an observable which pertains to an {\sl internal}
degree of freedom of the particle, so that it is difficult to conceive a {\sl direct}
measurement of $S_z$. However, let us consider the elementary observable
$T_{up}$ whose outcome 1 {\sl localizes} the atom in the upper exit of the magnet. 
According to the laws of quantum mechanics, $T_{up}$ is perfectly correlated
with $S_z$: the particle goes out through the upper exit
if and only if the outcome of a measurement of $S_z$ is $s_z=1$;
hence,
$S_z$ is assigned the value $1$ when the out-coming particle is
localized in the upper exit by a measurement of the localization observable $T_{up}$.
\par
In this experiment the assignment of the spin's vale, $1$ or $0$, is an evaluation by $T_{up}$.
\begin{figure}[ht]
\begin{center}
\includegraphics[width=0.5\textwidth]{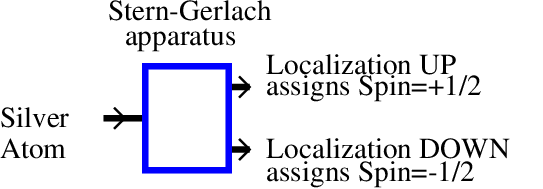}
\caption{Spin's value assigned by localization measurement}
\end{center}
\end{figure}
\vskip.5pc\noindent
In the following example a simultaneous value assignment to {\sl non-comeasurable observables} is realized by evaluations.
\vskip.5pc\noindent
{\bf Example 2.2.}
We consider a typical double-slit experiment (Fig. 2), where outcome 1 of the measurement of
the elementary observable $Q_S$ localizes 
the particle in the upper slit at the time of the crossing of the screen supporting the slits.
$Q_S$ {\it does not commute} with the projection
$Q_F$ whose outcome 1 localizes the particle in a fixed region on the final screen,
because $Q_S$ and $Q_F$ are positions at different times.
Therefore, the measurement of $Q_S$ is {\sl in principle}
forbidden for a particle whose final localization is measured by $Q_F$.
However, under suitable conditions \cite{Nistico}, an observable $T_S$ exists
such that $[T_S,Q_F]={\bf 0}$, whose outcomes are
perfectly correlated with the outcomes of $Q_S$ in every
simultaneous measurement of $T_S$ and $Q_S$. So, by measuring
$T_S$ and $Q_F$ together,
$Q_S$ is evaluated together to the direct measurement of $Q_F$, also if 
$Q_S$ and $Q_F$ canot be measured together. 
\begin{figure}[ht]
\begin{center}
\includegraphics[width=0.4\textwidth]{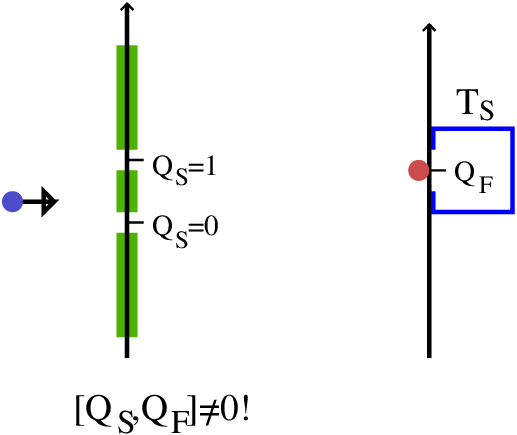}
\caption{Which Slit $Q_S$ is `evaluated' by measuring $T_S$}
\end{center}
\end{figure}
\vfill\eject\noindent
\subsection{Consistent Value Assignment by Evaluation}
The concept of {\sl consistent value assignment} can be now formally introduced. Let $A$ be a quantum observable. As already argued,
if $A$ is actually measured, then the value assignment to $A$ of the actual outcome is
consistent with all actually occurring phenomena, i.e. with all
occurrences of actual measurement outcomes of other observables $B$; then, the domain of the observables $B$
for which consistency is guaranteed with their measurent' outcomes is the commutant 
$\{A\}'=\{B\in\Omega({\mathcal H})\mid [B,A]=\nop\}$.
Since
the value assignment to $A$ does not necessarily originate from a direct measurement of $A$,
as in the case of Examples 2.1 and 2.2,
the following general
concept of consistent value assignment can be formulated.
\vskip.5pc\noindent{\bf Definition 2.3.}
{\sl Let $A$ be a quantum observable. A value assignment to $A$ is said to be consistent
with respect to the quantum state $\rho$ of the system,
whenever the following statements hold.
\begin{itemize}
\item[(V.1)]
If $A$ is actually measured, then its outcome coincides with the assigned value.
\item[(V.2)]
If $B$ is actually measured together with the value assignment to $A$, then every correlation that holds between
the outcomes of $B$ and $A$ whenever both measured in the state $\rho$, is identically
satisfied by the outcome of $B$ and the value assigned to $A$.
\end{itemize}}
\noindent
These conditions guarantee that the value assigned to $A$ in correspondence with a concrete specimen
can be treated in all respects as {\sl objective}, without the risk
of logical or empirical inconsistencies,
also if it is not the outcome of an actual measurement of $A$,
when applied to that specimen.
\par
Since in quantum theory all correlations between the outcomes of two observables $A$ and $B$
trace back to correlations, ruled over by Theorem 2.2, between all pairs of elementary observables $E^A_{\lambda_2}-E^A_{\lambda_1}$,
$E^B_{\mu_2}-E^B_{\mu_1}$,
conditions (V.1), (V.2) can be equivalently re-formulated in terms of elementary observables.
Namely, given the elementary observable $E$ recipient of the value assignment, let $F$ be another
elementary observable that can be measured on the specimen the value assignment for $E$
is referred to, let ${\mathcal F}_0$ be the set of all such elementary observables.
Once denoted by $p_\rho(E\& F)$ the probability that both the value assigned to $E$ and the
outcome for $F$ are 1, taking into account also Remark 2.1, the conditions (V.1), (V.2) become
\begin{itemize}
\item[{\rm (C.1)}]
If $F\in{\mathcal F}_0$ and $[ F, E]={\bf 0}$ then $p_\rho(E\& F)=Tr(\rho E F)$ and $p_\rho(E'\& F)=Tr(\rho E' F)$;
\item[{\rm (C.2)}]
if $\{F_j\}_{j\in J}\subseteq {\mathcal F}_0$ is any countable
family such that $\sum_{j\in J} F_j\equiv  F\in{\mathcal
F}_0$,
\item[]
then \quad$p_\rho( E\& F)=\sum_{j\in J}p_\rho( E\&
F_j)$ and $p_\rho( E'\& F)=\sum_{j\in J}p_\rho( E'\& F_j)$.
\end{itemize}\vskip.5pc\noindent
Now we can prove that the value assignment provided by evaluation is a consistent value assignment.
Let us consider an elementary observable $E\in\Pi({\mathcal H})$ that admits an evaluation by $T$ with respect to $\rho$,
i.e. such that $E\quad^\rho\lrcorr T$. Let $E$ be assigned the value obtained as outcome of a measurement of $T$.
This value assignment requires an actual measurement of $T$,
therefore the domain ${\mathcal F}_0$ where consistency can be checked is
${\mathcal F}_0=\{F\in\Pi({\mathcal H})\mid [T,F]=\nop\}\equiv{\mathcal F}(T)$.
Notice that $E\in{\mathcal F}(T)$ holds, of course.
The following theorem is proved in \cite{Ev}
\vskip.5pc\noindent
{\bf Theorem 2.3.}
{\sl
Fixed $E\in{\mathcal F}(T)$, for every quantum state
$\rho$ the mappings
\par
$h_\rho(E\&\,\cdot\,):{\mathcal F}(T)\to[0,1]$, $h_\rho(E\& F)=Tr(\rho E F E)$, and\par
$h_\rho(E'\&\,\cdot\,):{\mathcal F}(T)\to[0,1]$, $h_\rho(E'\& F)=Tr(\rho E' F E')$
\par\noindent
are the unique functionals that satisfy the conditions (C.1), (C.2)}
\vskip.5pc\noindent
The following simple theorem accomplishes our task.
\vskip.5pc\noindent
{\bf Theorem 2.4.}
{\sl Given two elementary observables $E,T\in\Pi({\mathcal H})$, if $E\quad^\rho\lrcorr T$,
so that $T$ can evaluate $E$,
then for all $F\in{\mathcal F}(T)$ the equalities
$$h_\rho(E\& F)=Tr(\rho T F)\;\hbox{ and }\;h_\rho(E'\& F)=Tr(\rho T' F)$$
hold.
Therefore, $h_\rho(E\& F)=Tr(\rho T F)$ satisfies (C.1) and (C.2).
}
\vskip.5pc\noindent
{\bf Proof.}
If $E\quad^\rho\lrcorr T$, then $ E\rho= T\rho$ and $\rho E=\rho  T$ hold;
therefore $h_\rho (E\& F)=Tr(\rho E F E)=Tr( E\rho E F)=Tr( T\rho E F)
=Tr( T\rho T F)=Tr(\rho T F)$ holds because $[ F, T]={\bf 0}$. Analogously, since
$E\quad^\rho\lrcorr T$ implies $E'\quad^\rho\lrcorr T'$, the equality $h_\rho(E'\& F)=Tr(\rho T' F)$ is proved.
\vskip.5pc\noindent
Thus, we can state the following theorem.
\vskip.5pc\noindent
{\bf Theorem 2.5.}
{\sl Let $A,T$ be quantum observables such that $A\quad^\rho\lrcorr T$.
The evaluation of $A$ by $T$ is a consistent value assignment and $p_\rho=h_\rho$ on ${\mathcal F}(T)$.}
\vskip.5pc\noindent
{\bf Remark 2.2.}
While if a value assignment to $E$ is ruled over by probabilities satisfying
conditions (C.1) ans (C.2) then its consistency is ensured, the necessary form
$Tr(\rho EFE)\equiv h_\rho(E\& F)$ of these probabilities does not always
corresponds to the probability of a value assignment to $E$.
Indeed, the probability $p_\rho(E\& F)$ of {\sl any} value assignment always satisfies
$$
p_\rho(E'\& F)+p_\rho(E\& F)=Tr(\rho F)\hbox{ for all }F\in{\mathcal F}_0,\leqno(C.3)
$$
independently of its consistency.
Instead, in general
\begin{description}
\item[$Tr(\rho F)$] $=Tr(\rho[ E+ E'] F[ E+\hat E'])
=Tr(\rho EFE)+Tr(\rho E'FE')+2{\bf Re}Tr(\rho E F E')$
\item[\qquad] $=h_\rho(E\& F)+h_\rho(E'\& F)+2{\bf Re}Tr(\rho E F E')$.
\end{description}
Therefore, if $[ F, E]\neq{\bf 0}$ the presence of a non-vanishing {\sl interference} term
$2{\bf Re}Tr(\rho E F E')$ cannot be excluded, and (C.3) is violated.
In this case no value assignment to $E$ can be ruled over by $h_\rho$.
On the contrary, the value assignment by evaluation naturally satisfies (C.3). Indeed,
\begin{description}
\item[$h_\rho(E\& F)+h_\rho(E'\& F)$]
$=Tr(\rho EFE)+Tr(\rho E'FE')=Tr(E\rho EF)+Tr(E'\rho E'F)$
\item[\qquad\qquad\qquad\qquad]$=Tr(T\rho EF)+Tr(T'\rho E'F)
=Tr(T\rho TF)+Tr(T'\rho T'F)$
\item[\qquad\qquad\qquad\qquad]
$=Tr(T\rho FT)+Tr(T'\rho FT')=
Tr(\rho T F)+Tr(\rho T' F)=Tr(\rho F)$.
\end{description}
The consistency of the value assignment by evaluation, proved for elementary observables,
extends to general observables. Let $A$ be an observable that can be evaluated by $S$
with respect to the quantum state $\rho$:  $A\quad^\rho\lrcorr S$. The value assignment to
$A$  through evaluation by $S$ requires an actual measurement of $S$; therefore, the domain
where the consistency must be checked is the commutant $\{S\}'$. It is just a matter of mathematical
computation based on theorem 2.4 to prove the following theorem.
\vskip.5pc\noindent
{\bf Theorem 2.6.}
{\sl If $A\quad^\rho\lrcorr T$, then the value assignment to $A$ through evaluation by $T$
satisfies conditions (C.1) and (C.2).
Thus, it is a consistent value assignment.}
\subsection{Key attainments}
It is important to underline the following remarks,
which are valid whenever $A\quad^\rho\lrcorr T$.
\begin{description}
\item[{\rm (E.1)}]
To assign $A$ the value evaluated by $T$
is consistent with all outcomes of actually occurring phenomena,
i.e. with all actually performed measurements of observables $B$ from $\{T\}'$, also if $[B,A]\neq{\bf 0}$.
In other words,
the value assigned to $A$ by evaluation can be treated as an objective value without the risk
of logical or empirical inconsistencies, for each concrete specimen the eavaluation refers to.
\item[{\rm (E.2)}]
The consistency of assigning $A$ the outcome of an evaluating observable $T$
is guaranteed, according to the Quantum Theory presented in section 2.1, if $T$ {\sl is actually measured};
e.g., consistency is {\sl not ensured} if $T$ is only evaluated.
\item[{\rm (E.3)}]
 It is possible
to exploit evaluations for consistently assigning non-commuting
observables simultaneous values. Indeed, let us consider two elementary observables $E_1$, $E_2$ such that
$E_1\quad^\rho\lrcorr T_1$ and $E_2\quad^\rho\lrcorr T_2$,
so that they can be evaluated by $T_1$ and $T_2$, respectively, in the state $\rho$.
If $[T_1,T_2]={\bf 0}$ and $E_1,E_2\in{\mathcal F}(T_1)\cap{\mathcal F}(T_2)$,
the evaluations of both $E_1$ and $E_2$ can be
obtained simultaneously by measuring together $T_1$ and $T_2$. Therefore $E_1$ and $E_2$ can be consistently assigned
simultaneous values, no matter if $[E_1, E_2]\neq{\bf 0}$.
\end{description}
\section{Emergence of classicality explained through evaluations}
The impossibility of a
consistent value assignment to the observables involved in some physical phenomena
makes meaningless the existence itself of the physical property
corresponding to a non-assignable value, and therefore
a classical explanation of these phenomena is not expected to exist. 
As a consequence, in general classicality does not emerge if a consistent value assignment
does not exist: non-classicality originates from this impossibilty.
\par
In fact, emergence of classicality is observed in physical phenomena 
for physical systems characterized by ``classical''
features, in relation to the ``classical'' observables.
The present section is devoted to investigate the possibility that emergence of
classicality can be derived completely within quantum theory itself, by a quantum
theoretical mechanism that allow for a consistent value assignment to non-commuting
observables corresponding to the magnitudes that classically describe the behavior of the system.
\par
As an archetypical phenomenon for a physical system with classical features we consider
the motion of a ``rigid body''. In this case two non commuting observables that correspond to the
magnitudes that characterize the motion from a classical point of view are the
position ${\bf Q}_{CM}$ of the center of mass and its velocity
${\bf V}_{CM}=\frac{d{\bf Q}}{dt}=\dot{\bf Q}$.
Though $[{\bf Q}_{CM},{\bf V}_{CM}]\neq\nop$, we shall show how a rigidity condition allow for
a consistent value assignment to both ${\bf Q}_{CM}$ and ${\bf Q}_{CM}$, explaining emergence
of classicality for the motion of the center of mass.
\subsection{Toy case of rigid body}
Let us consider a physical system composed by four identical but distinguishable particles with the same mass $\mu$, e.g. atoms, so that
the Hilbert space of its quantum theory is
${\mathcal H}=L_2(\RR^{4\cdot 3})$, isomorphic to ${\mathcal H}_1\otimes{\mathcal H}_2\otimes{\mathcal H}_3\otimes{\mathcal H}_4$
where each ${\mathcal H}_j=L_2(\RR^3)$  is the Hilbert space of the quantum theory of the $j$-th particle as single particle.
The operator that represents the position of particle $j$ is the multiplication operator:
${\bf Q}_j\Psi({\bf x}^{(1)},{\bf x}^{(2)},\dots)={\bf x}^{(j)}\Psi({\bf x}^{(1)},{\bf x}^{(2)},\dots)$.
Accordingly, $[{\bf Q}_j,{\bf Q}_k]=\nop$, i.e. the four positions can in principle be measured together.
We restrict to the case where the interaction undergone by the particles depends on their positions,
i.e. $H=-\frac{1}{2\mu}\sum_j(\nabla_{{\bf x}^{(j)}}^2+U({\bf x}^{(1)},{\bf x}^{(2)},\dots)$.
Accordingly,
the operator that represent the velocity of particle $j$ turns out to be
$${\bf V}_j=i[H,{\bf Q}_j]=-\frac{i}{\mu}\nabla_{{\bf x}_j},\hbox{ so that } 
[{\bf V}_j,{\bf V}_k]=\nop,\quad [{\bf Q}_j,{\bf V}_k]=\frac{i}{m}\delta_{jk}.\eqno(3.1)$$
Now we introduce a {\sl rigidity} condition for the state vector, which characterizes the system as a rigid body:
the distances between the particles of any pair is constant in time.
To theoretically state this rigidity condition the operator ``distance between particles $j$ and $k$'' is introduced as
$D_{jk}=\sqrt{({\bf Q}j-{\bf Q}_k)^2}$. Then the system is a rigid body if its state vector $\psi^{rig}_t$
is such that every measurement of of the positions always yields the same value $d_{jk}$ 
for the distances $D_{jk}$ and at any time.
Therefore, the following eingenvalue equations must hold for all $t$.
$$D_{jk}\Psi^{rig}_t=d_{jk}\Psi^{rig}_t,\quad \forall j,k,t\,.\eqno(3.2)$$
From the point of view of classical phyisics two magnitudes that describe the motion
of the body are the position of the center of mass and its velocity.
Now, in the quantum theory of this system the position of the center of mass is represented by
$${\bf Q}_{CM}=\frac{1}{4}({\bf Q}_1+{\bf Q}_2+{\bf Q}_3+{\bf Q}_4).\eqno(3.3)$$
The velocity of the center of mass is represented by the operator
$${\bf V}_{CM}=\dot{\bf Q}_{CM}=i[H,{\bf Q}_{CM}]
=\frac{1}{4}(\dot{\bf Q}_1+\dot{\bf Q}_2+\dot{\bf Q}_3+\dot{\bf Q}_4).$$
By (3.1) we derive
$[{\bf Q}_{CM},{\bf V}_{CM}]\neq\nop$, so that the position and the velocity of the center of mass
are not measurable together.
We now show that the rigidity condition (3.2) allow for a simultaneous consistent value assignment to
${\bf Q}_{CM},{\bf V}_{CM}$ by evaluation, and then a classical description of the motion of the 
center of mass is possible for the assigned values.
Let us suppose that $d_{12}=1$, $d_{23}=1$, $d_{34}=1$, $d_{14}=3$ hold,
so that the positions of the four atoms, whenever measured, are rigidly located on a segment of length 3, 
at the extremals of intervals of length 1. This implies that 
$${\bf Q}_{CM}\Psi^{rig}_t=\frac{1}{2}({\bf Q}_1+{\bf Q}_4)\Psi^{rig}_t=\frac{1}{2}({\bf Q}_2+{\bf Q}_3)\Psi^{rig}_t.\eqno(3.4)$$
The general Schroedinger equation $i\frac{\partial}{\partial t}\Psi_t=H\Psi_t$ and equation (3.4) impliy
\begin{description}
\item[${\bf Q}_{CM}H\Psi^{rig}_t$]$={\bf Q}_{CM}i\frac{\partial}{\partial t}\Psi^{rig}_t=i\frac{\partial}{\partial t}{\bf Q}_{CM}\Psi^{rig}_t=
i\frac{\partial}{\partial t}
\frac{1}{2}({\bf Q}_1+{\bf Q}_4)\Psi^{rig}_t$
\item[\qquad\qquad]$=\frac{1}{2}({\bf Q}_1+{\bf Q}_4)i\frac{\partial}{\partial t}\Psi^{rig}_t=
\frac{1}{2}({\bf Q}_1+{\bf Q}_4)H\Psi^{rig}_t$.
\end{description}
This equation and (3.4) imply
\begin{description}
\item[${\bf V}_{CM}\psi_t^{rig}$]$=i[H,{\bf Q}_{CM}]\psi_t^{rig}=iH\frac{1}{2}({\bf Q}_1+{\bf Q}_4)\Psi^{rig}_t
-i\frac{1}{2}({\bf Q}_1+{\bf Q}_4)H\Psi^{rig}_t$\hfill{(3.5)}
\item[\qquad]$=i[H,\frac{1}{2}({\bf Q}_1+{\bf Q}_4)]\psi^{rig}_t=
\frac{1}{2}({\bf V}_1+{\bf V}_4)\Psi^{rig}_t$.
\end{description}
\par\noindent
Let us define ${\bf T}_Q=\frac{1}{2}({\bf Q}_2+{\bf Q}_3)$ and ${\bf T}_V=\frac{1}{2}({\bf V}_1+{\bf V}_4)$.
Equations (3.4) and (3.5) impy
$$ {\bf T}_Q\Psi^{rig}_t = {\bf Q}_{CM}  \Psi^{rig}_t,\quad {\bf T}_V \Psi^{rig}_t={\bf V}_{CM} \Psi^{rig}_t.$$
Hence 
${\bf T}_Q$ and ${\bf T}_V $ evaluate ${\bf Q}_{CM}$ and ${\bf V}_{CM}$ respectively.
Moreover $[{\bf T}_V,{\bf T}_Q]=\nop$.
Therefore, by (E.3) in section 2.4, though the position and the velocity of the center of mass cannot be measured together,
they admit a simultaneous consistent value assignment.
Thus classicality holds for the motion of the center of mass.
\subsection{More Realistic Rigid Body}
The model of rigid body of the previous subsection is not realistic, because it is not the case that the position of
some particles is so exactly related to the position of the other particles in quantum theory.
In particular equation (3.2) hardly can be satisfied by a $\Psi_t^{rig}\in L_2(\RR^{12})$.
Now the problem is faced with a more realistic model of an homogeneous solid body
composed by a very large number $N$ of identical but distinguishable particles of mass  $\mu$.
The Hilbert space of its quantum theory is
${\mathcal H}=L_2(\RR^{3N})$ isomorphic to ${\mathcal H}_1\otimes{\mathcal H}_2\otimes\cdots\otimes{\mathcal H}_N$,
where each ${\mathcal H}_j=L_2(\RR^3)$ is the Hilbert space of the quantum theory of the $j$-th particle as single particle.
By restricting to interactions that depend on the positions, we can re-obtain (3.1). 
The position of the center of mass is represented by the operator ${\bf Q}_{CM}=\sum_n^N\frac{{\bf Q}_n}{N}$.
If the measured value of each particle is ${\bf q}_n$, then the measured value of ${\bf Q}_{CM}$ 
is  ${\bf q}_{CM}=\sum_n^N\frac{{\bf q}_n}{N}$.
\par
The operator that represents the velocity of the center of mass is ${\bf V}_{CM}=i[H,{\bf Q}_{CM}]=\sum_n^N\frac{\dot{\bf Q}_n}{N}$,
where $[\dot{\bf Q}_j,\dot{\bf Q}_k]=\nop$.
However
$[{\bf Q}_{CM},{\bf V}_{CM}]\neq\nop$, so that  the position and the velocity of the center of mass cannot be measured together.
\par
Each measured position ${\bf q}_n$ has not a determined value,  because it is ruled over by
a probability density $\rho_n({\bf x})$ such that the expected value $\langle {\bf q}_n\rangle=\int{\bf x}\rho_n({\bf x})d{\bf x}^3$ is determined.
Now ${\bf q}_{CM}=\sum_n^N\frac{{\bf q}_n}{N}=\sum_n^N\frac{{\bf q}_n-\langle {\bf q}_n\rangle}{N}+\sum_n^N\frac{\langle {\bf q}_n\rangle}{N} $.
The probability density $\rho_n({\bf x})$ is nearly symmetric around $\langle{\bf q}_n\rangle$. So,
since the number $N$ is very large, we can conclude $\sum_n^N\frac{{\bf q}_n-\langle {\bf q}_n\rangle}{N}=0$, i.e.
$${\bf q}_{CM}=\sum_n^N\frac{\langle {\bf q}_n\rangle}{N}. $$
\begin{figure}[ht]
\begin{center}
\includegraphics[width=0.5\textwidth]{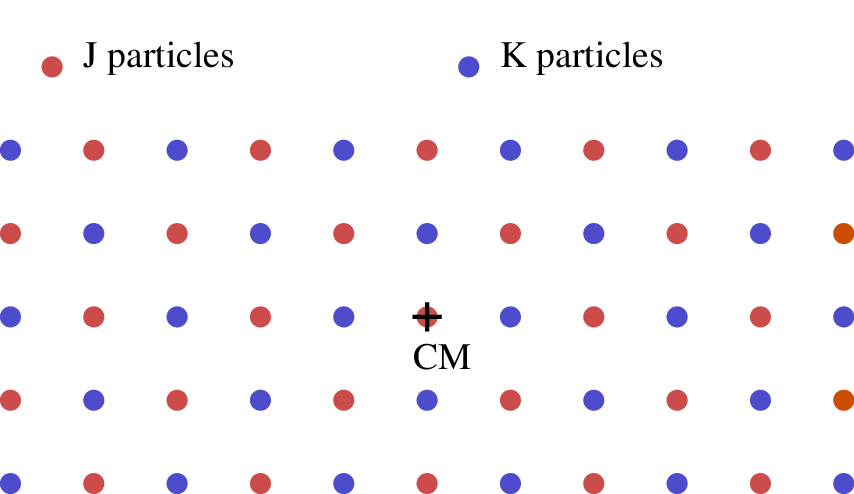}
\caption{${\bf q}_{CM}={\bf q}_{CM}^J={\bf q}_{CM}^J$}
\end{center}
\end{figure}
A rigidity condition for the homogeneous body can be formulated by stating that the 
{\sl center of mass can be equivalently
determined with different suitable subsets of particles}, for instance the subsets with positions
$\{{\bf q}_j\mid j\in J\}$ and $\{{\bf q}_k\mid j\in K\}$  where $J\cap K=\emptyset$ (see fig. 3). 
Their centers of mass are represented by the operators
${\bf Q}^J_{CM}=\sum_{j\in J}\frac{{\bf Q}_j}{N_J}$ and ${\bf Q}^K_{CM}=\sum_{k\in K}\frac{{\bf Q}_k}{N_K}$.
So, if the state vector of the body is  state of rigidity $\Psi_t^{rig}$, the rigidity condition 
$${\bf q}_{CM}=\sum_n^N\frac{\langle{\bf q}_n\rangle}{N}=\sum_{j\in J}\frac{{\bf q}_j}{N_J}=
\sum_{j\in J}\frac{\langle{\bf q}_j\rangle}{N_J}\equiv {\bf q}_{CM}^J=
\sum_{k\in K}\frac{{\bf q}_k}{N_K}=\sum_{k\in K}\frac{\langle{\bf q}_k\rangle}{N_K}\equiv{\bf q}_{CM}^K\eqno(3.6)$$
implies that ${\bf Q}_{CM}$, ${\bf Q}_{CM}^J$, ${\bf Q}_{CM}^K$ are perfectly correlated. Therefore
the following equations must hold. 
$${\bf Q}_{CM}\Psi_t^{rig}={\bf Q}^J_{CM}\Psi_t^{rig}=\sum_{j\in J}\frac{{\bf Q}_j}{N_J}\Psi_t^{rig}=
{\bf Q}^K_{CM}\Psi_t^{rig}=\sum_{k\in K}\frac{{\bf Q}_k}{N_K}\Psi_t^{rig}.\eqno(3.7)$$
Following the same derivation to obtain (3.5), from (3.7) we can deduce
$${\bf Q}_{CM}H\Psi_t^{rig}={\bf Q}^J_{CM}H\Psi_t^{rig}=
{\bf Q}^K_{CM}H\Psi_t^{rig}.\eqno(3.8)$$
Once defined ${\bf T}_V=i[H,{\bf Q}_{CM}^K]=\sum_{k\in K}\frac{\dot{\bf Q}_k}{N_K}$ and ${\bf T}_Q={\bf Q}_{CM}^J=\sum_{j\in J}\frac{{\bf Q}_j}{N_J}$,
(3.7), (3.8) rewrite as 
$${\bf T}_V\Psi_t^{rig}={\bf V}_{CM}\Psi_t^{rig}\hbox{ and }{\bf Q}_{CM}\Psi_t^{rig}={\bf T}_{CM}\Psi_t^{rig}.\eqno(3.9)$$
Since $=[{\bf T}_V,{\bf V}_{CM}]=[{\bf T}_Q,{\bf Q}_{CM}]=\nop$,
(3.9) implies that ${\bf Q}_{CM}\quad^\rho\lrcorr {\bf T}_Q$ and  ${\bf V}_{CM}\quad^\rho\lrcorr {\bf T}_V$; but $[{\bf T}_Q,{\bf T}_V]=\nop$;
thus ${\bf Q}_{CM}$ and ${\bf V}_{CM}$ can be simultaneously assigned consistent values by evaluation.
Thus, classicality holds for the motion of the center of mass also in this more realistic model.
\section{Appendices}
\subsection{A: Failure of classical theories}
We describe an experiment whose real performance led to contradict the following statement, necessary to every classical theory.
\begin{itemize}
\item[($\mathcal{CS}$)]
{\sl At any time every specimen of the physical system has a determined value $v({\mathcal A})$ of each of its magnitudes $\mathcal A$.
A measurement of $\mathcal A$, with outcome $v_0({\mathcal A})$, just
reveals this objective value.}
\end{itemize}
Namely, in this experiment any value assignment to the involved magnitudes is inconsistent
with the 
empirically validated locality principle and empirically validated relations among these values. 
\par
The physical system for this experiment is the couple $(P_L,P_R)$ of particles, couple emitted one at once
by a suitable source $S$ in such a way that after the emission the particles
$P_L$ and $P_R$
travel as free particles in opposite directions, say left h.s. and right h.s., respectively.
Once the particles are space-like separated they can 
undergo separated measurement procedures of 
a physical magnitude $\mathcal L$ on the left and of another physical magnitude 
$\mathcal R$ on the right.
The case can occur that for a particular pair $({\mathcal L},{\mathcal R})$ of these magnitudes
a setting of the source exists such that the outcomes $v_0({\mathcal L})$ and $v_0({\mathcal R})$
obtained by actual measurements are constrained to each other, for instance
in such a way that the correlation 
$$
\hbox{ if}\quad v_0({\mathcal L})=1\quad\hbox{then}\quad v_0({\mathcal R})=1\leqno(co.0)
$$
holds whenever $\mathcal L$ and $\mathcal R$ are actually measured on a couple emitted in that setting.
\par
Now we derive a general implication of statement ($\mathcal{CS}$) when the correlation (co.0) holds.
Let a measurement of $\mathcal L$ yield the outcome $v_0({\mathcal L})=1$.
If also $\mathcal R$ is measured then (co.0) implies $v_0({\mathcal R})=1$. 
The {\sl objective} value $v({\mathcal R})$ assumed by (${\mathcal CS}$) 
cannot depend on the choice of measuring $\mathcal R$ or not,
because $P_L$ and $P_R$ are space-like separated. 
Therefore,
\vskip.5pc\noindent
{\bf Proposition 4.1.}
{\sl
If ($\mathcal{CS}$) holds together with the locality principle, then (co.0) implies
$$
\hbox{ if}\quad v_0({\mathcal L})=1\quad\hbox{then}\quad v({\mathcal R})=1\;.\eqno(4.1)
$$}
In fact, more than one setting has been concretely found \cite{Torgerson}-\cite{Barbieri} such that the following correlations hold for three particular pairs
$({\mathcal L}_1,{\mathcal R}_1)$,  $({\mathcal L}_2,{\mathcal R}_1)$,  $({\mathcal L}_2,{\mathcal R}_2)$ of magnitudes:
\begin{itemize}
\item[(co.1)]
whenever ${\mathcal L}_1$ and ${\mathcal R}_1$ are measured then $v_0({\mathcal L}_1)=1$ implies 
$v_0({\mathcal R}_1)=1$;
\item[(co.2)]
whenever ${\mathcal L}_2$ and ${\mathcal R}_1$ are measured then $v_0({\mathcal R}_1)=1$ implies 
$v_0({\mathcal L}_2)=1$;
\item[(co.3)]
whenever ${\mathcal L}_2$ and ${\mathcal R}_2$ are measured then $v_0({\mathcal L}_2)=1$ implies  
$v_0({\mathcal R}_2)=1$.
\end{itemize}
These correlations turned out to be indisputable facts in that setting.
\vskip.4pc\noindent
Let us now suppose that only ${\mathcal L}_1,{\mathcal R}_2$ are actually measured.
If ($\mathcal{CS}$) is assumed to hold, we can make use of Prop. 4.1 and correlation (co.1) 
so that 
$$
v_0({\mathcal L}_1)=1\quad\hbox{implies}\quad v({\mathcal R}_1)=1
$$
holds.
Hence Prop. 4.1 would imply $v({\mathcal L}_2)=1$ by (co.2) if ${\mathcal R}_1$ were measured, because in this case
$v({\mathcal R}_1)=v_0({\mathcal R}_1)=1$. But according to the locality principle the objective value $v({\mathcal L}_2)$ does not depend on the choice of measuring 
${\mathcal R}_1$ or not, because $P_L$ and $P_R$ are space-like separated. Therfeore
$$
\hbox{ if}\quad v_0({\mathcal L}_1)=1\quad\hbox{then}\quad v({\mathcal L}_2)=1\eqno(4.2)
$$
holds if ($\mathcal{CS}$) holds. Iterating this argument we infer the implication
$$
\hbox{ if}\quad v_0({\mathcal L}_1)=1\quad\hbox{then}\quad v({\mathcal R}_2)=1\;.
$$
Thus,  ($\mathcal{CS}$) and (co.1-3) imply
$$
\hbox{ if}\quad v_0({\mathcal L}_1)=1\quad\hbox{then}\quad v_0({\mathcal R}_2)=1\;.\eqno(4.3)
$$
In fact, a setting
has been concretely found
for which (co.1-3) hold, but in some actually performed measurements of 
$({\mathcal L}_1,{\mathcal R}_2)$ the outcome turned to be
$$
v_0({\mathcal L}_1)=1\quad\hbox{and}\quad v_0({\mathcal R}_2)=0,\leqno{(\mathcal{ER})}
$$
in contradiction with statement (4.3).
Since the validity of correlatiions (co.1-3) is ascertained, statement $({\mathcal CS})$
turns out to e empirically inconsistent.
\vskip.4pc
This experiment is essentially the experiment conceived by Hardy \cite{Hardy},
actually realized in equivalent forms
\cite{Torgerson}--\cite{Barbieri}.
Hardy's aim was also to contribute to the debate about locality of quantum physics. 
According to our analysis here, the results of its performance lead to conclude that statement
($\mathcal{CS}$) is invalid, while the locality principle is not affected.
\subsection{B: Proof of theorem 2.2}
Theorem 2.2 is an extension of the following lemma, whose proof can be easily
obtained from Prop. 2.1 in \cite{Ev}.
\vskip.5pc\noindent
{\bf Lemma 4.1.} {\sl Two elementary observables $E,T\in\Pi({\mathcal H})$ are perfectly correlated
with respect to the quantum state $\rho$,
i.e. $E\quad^\rho\lrcorr T$, if and only if
\begin{itemize}
\item[i)]
$[E,T]=\nop$,
\item[ii)]
$E\rho=T\rho$ or equivalently $\rho T=\rho E$.
\end{itemize}
}
\vskip.5pc\noindent
{\bf Theorem 2.2.}
{\sl Two quantum observables $A$ and $T$ are perfectly correlated with respect to the quantum state $\rho$,
in formula $A\quad^\rho\lrcorr T$, if and only if
\begin{itemize}
\item[i)]
$[A,T]=\nop$,
\item[ii)]
$A\rho=T\rho$ or $\rho T=\rho A$.
\end{itemize}
}
\vskip.4pc\noindent
{\bf Proof.}
If $A$ and $T$ are perfectly correlated with respect to $\rho$, then 
$[(E^A_{\lambda_2}-E^A_{\lambda_1}),(E^T_{\lambda_2}-E^T_{\lambda_1})]=\nop$ holds for all $\lambda_1,\lambda_2$,
since (i) implies $[E^A_\lambda,E^T_\mu]=\nop$ \cite{Kreyszig}.
Hence, according to Lemma 4.1 the equalities
$Tr\{\rho(E^A_{\lambda_2}-E^A_{\lambda_1})(E^T_{\lambda_2}-E^T_{\lambda_1})\}=
Tr\{\rho(E^A_{\lambda_2}-E^A_{\lambda_1})\}
=Tr\{\rho(E^T_{\lambda_2}-E^T_{\lambda_1})\}$
hold, which imply
$$
Tr\{\rho(E^A_{\lambda_2}-E^A_{\lambda_1})[\Id-(E^T_{\lambda_2}-E^T_{\lambda_1})]\}=
Tr\{\rho(E^T_{\lambda_2}-E^T_{\lambda_1})[\Id-(E^A_{\lambda_2}-E^A_{\lambda_1})]\}=0\,.\eqno(4.4)
$$
Now,
$(E^A_{\lambda_2}-E^A_{\lambda_1})[\Id-(E^T_{\lambda_2}-E^T_{\lambda_1})]$ and 
$(E^T_{\lambda_2}-E^T_{\lambda_1})[\Id-(E^A_{\lambda_2}-E^A_{\lambda_1})]$
are projection operators because $[A,T]=\nop$; then (4.4) implies
$$
\rho(E^A_{\lambda_2}-E^A_{\lambda_1})[\Id-(E^T_{\lambda_2}-E^T_{\lambda_1})]\}=
\rho(E^T_{\lambda_2}-E^T_{\lambda_1})[\Id-(E^A_{\lambda_2}-E^A_{\lambda_1})]=\nop,
$$, from which
$$
\rho(E^A_{\lambda_2}-E^A_{\lambda_1})=\rho(E^T_{\lambda_2}-E^T_{\lambda_1})\eqno(4.5)
$$
follows.
Now, by making use of spectral theory we find
$$
\rho A=\rho\int\lambda\, d_\lambda E^A_\lambda=\int\lambda\, \rho d_\lambda E^A_\lambda\quad\hbox{and}
\quad\rho T=\rho\int\lambda\, d_\lambda E^T_\lambda=\int\lambda\,\rho d_\lambda E^T_\lambda.\eqno(4.6)
$$
But (4.5) implies $\rho d_\lambda E^A_\lambda=\rho d_\lambda E^T_\lambda$, therefore
$$
\rho A=\int\lambda\, \rho d_\lambda E^A_\lambda=\int\lambda\,\rho d_\lambda E^T_\lambda=\rho T.
$$
\par
Now we have to prove the inverse implication, i.e. that
(i) $[A,T]=\nop$ and (ii) $\rho A=\rho T$ imply that $A$ and $T$ are perfectly correlated.
Now, $A$ and $T$ are perfectly correlated if and only if 
$(E^A_{\lambda_2}-E^A_{\lambda_1})$ and $(E^T_{\lambda_2}-E^T_{\lambda_1})$ do.
Therefore, in virtue of Lemma 4.1, it is sufficient to prove that
$$
[A,T]=\nop \hbox{ and } \rho A=\rho T\quad\hbox{imply\quad}
\rho(E^A_{\lambda_2}-E^A_{\lambda_1})=\rho(E^T_{\lambda_2}-E^T_{\lambda_1}),\; 
\forall \lambda_1,\lambda_2\,.\eqno(4.7) 
$$
Now, let $\chi_{(\lambda_1,\lambda_2]}$ be the characteristic functional of $(\lambda_1,\lambda_2]$;
according to the functional principle 
$$
(E^A_{\lambda_2}-E^A_{\lambda_1})=\chi_{(\lambda_1,\lambda_2]}(A)\quad\hbox{and}\quad
(E^T_{\lambda_2}-E^T_{\lambda_1})=\chi_{(\lambda_1,\lambda_2]}(T), \eqno(4.8)
$$
Given any self-adjoint operator $B$ a sequence of analytic functions 
$\{\varphi_n(\lambda)=\sum_ja^{(n)}_j\lambda^j\}$ exists such that 
$\langle\psi\mid \rho\chi_{(\lambda_1,\lambda_2]}(B)\psi\rangle=\lim_n\langle\psi\mid\rho \varphi_n(B)\psi\rangle$ for all $\psi$
in a dense domain.
Then 
$$
\langle\psi\mid \rho(E^A_{\lambda_2}-E^A_{\lambda_1})\psi\rangle=\lim_n\langle\psi\mid \rho\varphi_n(A)\psi\rangle=
\lim_n\langle\psi\mid \sum_ja^{(n)}_j\rho A^j\psi\rangle\quad\hbox{and}\eqno(4.9.i)
$$
$$
\langle\psi\mid \rho(E^T_{\lambda_2}-E^T_{\lambda_1})\psi\rangle=\lim_n\langle\psi\mid \sum_ja^{(n)}_j\rho T^j\psi\rangle.\eqno(4.9.ii)
$$
But 
$$
\rho A=\rho T\hbox{  and }  [A,T]=\nop\quad\hbox{imply}\quad \rho A^j=\rho T^j,\;\forall j.\eqno(4.10)
$$
By making use of (4.10) in (4.9.i,ii) we obtain 
$\langle\psi\mid \rho(E^A_{\lambda_2}-E^A_{\lambda_1})\psi\rangle=\langle\psi\mid \rho(E^T_{\lambda_2}-E^T_{\lambda_1})\psi\rangle$,
which implies (4.7).

\end{document}